\input{aipcheck}

\documentclass[
    ,final            
  ]
  {aipproc}
\newcommand{\chione}{\widetilde{\chi}_1^\pm}
\newcommand{\chionezero}{\widetilde{\chi}_1^0}
\newcommand{\chitwo}{\widetilde{\chi}_2^0}
\layoutstyle{6x9}

\begin{document}

\title{Distinguishing between SUSY and Littlest Higgs Model using trileptons at the LHC}

\classification{12.60.-i, 12.60.Jv}

\keywords      {Littlest Higgs Model, T-parity, Supersymmetry, LHC}

\author{Sudhir Kumar Gupta\footnote{skgupta@iastate.edu}}{
  address={Department of Physics, Iowa State University, Ames, IA 50011}
}

\begin{abstract}
Littlest Higgs model with T-parity and Minimal Supersymmtric standard Model (MSSM) with R-parity both give similar signatures in collider experiment with a huge amount of missing energy 
depending upon mass of the lightest T-odd/R-odd particle. In this talk, I will discuss possibility of distinguishing the two models at the LHC in hadronically quiet signal where masses of R-odd
particles are identical to masses of T-odd particles.
\end{abstract}

\maketitle

\section{Introduction}

Hints to much long awaited questions about the correct theory beyond the standard model (BSM) to describe new physics at the TeV scale is soon expected to reach us while the Large Hadron Collider 
(LHC) takes off for physics run. Once we detect new resonances, the next level goal will to identify the correct BSM model these belongs to. The reason being existence of several 
candidate theories sharing similar features.

The best known class of models which have close resemblance to each other include minimal Supersymmetric standard model (MSSM), Littlest Higgs Models and Universal extra dimension. These all advocate for a discrete $Z_2$ symmetry, under which  the new matter contents and their corresponding SM partners transform differently (In general new matter contents are odd under $Z_2$ and SM contents are even.) and thus provide a candidate for cold dark matter of the universe in the form of a lightest, neutral $Z_2$ odd particle (LOP) when it is conserved. The other phenomenological consequences include pair production of $Z_2$ odd partners and their cascade decays into LOPs which appear at the end of the cascades. 
 
At the LHC, confusion between two models can arise with models having particles with same spin SM partners,  similar couplings or even due to identical mass. The later possibility, which is major subject 
of the talk, is even more interesting as at the LHC, it is difficult to measure spin of the produced particles and  kinematic distributions will be identical in such a case. In this case, we need to fully rely upon the event rates to    distinguish between the two models vis a vis a precise information on the mass of the produced particles which can be translated to higher LHC Luminosities.


\section{The Littlest Higgs Model}

In little Higgs model~\cite{c7LHmin}, a global symmetry $SU(5)$ is
spontaneously broken down to $SO(5)$
at a scale $\Lambda \sim 4 \pi f \sim 10~\mbox{TeV}$. An $[SU(2) \otimes
U(1)]^2$ gauge symmetry is
imposed, which is simultaneously broken at $f$ to the diagonal subgroup
$SU(2)_L \otimes U(1)_Y$, which is identified with the SM gauge group.
Breaking of the $SU(5)$ leads to $14$ massless Goldstone bosons, which
transform under the $SU(2)_L \otimes U(1)_Y$ as a real singlet ($1_0$), a
real triplet ($3_0$), a complex doublet ($2_{\pm 1/2}$), and a
complex triplet ($3_{\pm1}$). The real singlet and the real triplet become
the longitudinal components of the gauge bosons associated with the broken
gauge groups, giving them masses of the order of $f$, while keeping the
complex doublet and the complex triplet massless. The
complex triplet and the neutral component of the complex doublet will
get mass of the order $f$ and $v$ (vev) respectively from a
Coleman-Weinberg Type potential which is induced by the gauge and Yukawa
couplings. The remaining three pseudo goldstone become the longitudinal
components of the SM gauge bosons.

In order to comply with strong constraints from electroweak precision data
on the Littlest Higgs model, one imposes  a discrete symmetry called 
T-parity. This maps the two pairs of gauge groups $SU(2)_i \otimes U(1)_i, i=1,2$
into each other. As a result of it, the corresponding gauge couplings become equal, with $g_1 = g_2$ and $g_1^\prime =g_2^\prime$.

Consequences of T-parity conservation in the the Little Higgs Models are same as that of R-parity in supersymmetry.

Mass spectrum of the Littlest Higgs Model with T-parity (LHT) can be essentially determined using three parameters: $f$, $k_l$ and $k_q$. Masses of T-odd heavy SM partners are given as follows: 

$m_{u_H} = \sqrt{2} \kappa_q f \left(1 - {v^2 \over 8 f^2} \right)$, $m_{d_H} = \sqrt{2} \kappa_q f$, $m_{l_H} = \sqrt{2} \kappa_l f \left(1 - {v^2 \over 8 f^2} \right)$, 
$m_{\nu_H} = \sqrt{2} \kappa_{\nu}f$, 
$m_{W_H} = m_{Z_H} = g f \left(1 - {v^2 \over 8 f^2} \right) \approx 0.65 f$,    
$m_{A_H} = {f g^\prime \over \sqrt{5}} \left(1 - {5 v^2 \over 8 f^2} \right)  \approx 0.16 f$.

Clearly, LHT spectrum can be essentially determined using three parameters: $f$, $k_l$ and $k_q$. In addition we have extra tops and extra Higgses whose spectrum can be determined using two additional parameters $\lambda_1$ and $\lambda_2$.

Electroweak precision data requires $\kappa \leq 4.8$ (for $f =1~\mbox{TeV}$). Also from the above equations, it is clear that $k_q > .11$ in order to avoid a charged dark matter. Negative searches of 
a heavy neutrino in direct dark matter detection has put a lower bound on $k_l > .11$~\cite{oursdm}.

\section{LHT-SUSY Distinction using hadronically quiet trileptons}
\begin{table}
\caption{LHT and SUSY  mass spectrum for fixed $f =500$ GeV and $\kappa_l,q =1$~\cite{ours}. All the masses are in GeV.}
\begin{tabular}{|rr@{.}lr@{.}lr@{.}lr@{.}lr@{.}lr@{.}l|cr@{.}lr@{.}lr@{.}l|}
\hline \hline 
\multicolumn{13}{|c|}{\bf LHT} & \multicolumn{7}{c|}{\bf SUSY} \\ \hline  
\multicolumn{1}{|c}{$f$} & \multicolumn{2}{c}{$m_{A_H}$} &
\multicolumn{2}{c}{$m_{Z_H}$} & \multicolumn{2}{c}{$m_{d_H}$} &
\multicolumn{2}{c}{$m_{u_H}$} & \multicolumn{2}{c}{$m_{l_H}$} &
\multicolumn{2}{c|}{$m_{\nu_H}$} & 
{\bf Case} & \multicolumn{2}{c}{$m_{\chionezero}$} &
\multicolumn{2}{c}{$m_{\chitwo}$} & \multicolumn{2}{c|}{$m_{\chione}$} \\  
\hline 
500 & 66 & 2 & 316 & 7 & 707 & 1 & 685 & 7 & 707 & 1 & 685 & 7 & 
{\bf SS1} & 65 & 9 & 314 & 9 & 314 & 9 \\ 
\multicolumn{13}{|c|}{} & {\bf SS2} & 63 & 7 & 314 & 9 & 318 & 1 \\ 
\hline \hline
\end{tabular} 
\label{mass_table}
\end{table}

In the LHT, hadronically quiet trileptons can be produced via $q\bar{q^{\prime}} \to W_H^{\pm} Z_H$,  where,  $W_H^{\pm} \to A_H W^{\pm} \to A_H l^{\prime \pm} \nu_{l^{\prime}}$ and $Z_H
\to A_H Z \to A_H l^{\pm} l^{\mp}$.  

The aforementioned signature in SUSY can arise due to $q\bar{q^{\prime}} \to \chione \chitwo$ production  with $\chione \to\nu_{l^{\prime}} {\widetilde l^{\prime
\pm}} \to \nu_{l^{\prime}} {l^{\prime \pm}} \chionezero$ and $\chitwo \to
l^{\pm} {\widetilde l^{\mp}} \to l^{\pm} l^{\mp} \chionezero $.

We generated events using CalcHEP~\cite{calchep} and later interfaced these to PYTHIA~\cite{pythia} for analysis purpose. The major SM  
 background is due to $WZ$ production. We have also incorporated subdominant backgrounds due to top-pair as well as due to tri-gauge boson productions.

For the event analysis, we implemented following additional cuts apart from the basics cuts in order to suppress the background,
\begin{itemize}
\item 
Missing transverse energy cut, $E_T{\!\!\!\!\!\!/\ } \geq 100$ GeV.
\item 
$m_{l^+l^-} > 20$ GeV to ensure absence of leptons emitted from off-shell photons.  
\item
$|m_{l^+l^-} - m_Z | > 15$ GeV and $|m_T{(l E_T{\!\!\!\!\!\!/\ })} - m_W| > 15$ GeV to reject leptons arising due to 
$Z$ and $W$ background respectively.
\end{itemize}

In order to match LHT spectrum with SUSY, we equate squarks and sleptons masses to those of the heavy quarks and leptons. To match masses of heavy gauge bosons to those of neutralinos and charginos we set 
$m_1 = m_{A_H}$, $m_2 =  m_{Z_H}$ for $m_2 > \mu$ or $\mu =  m_{Z_H}$ for $\mu > m_2$ which correspond to SUSY Scenario 1 and 2 (SS1 and SS2) respectively. Also, we set $m_3 = 5$ TeV to decouple the gluino. The other SUSY parameters are $\tan\beta =10$,  $m_A =850$ GeV and $m_h =120$ GeV. 

\begin{figure}
      \resizebox{75mm}{!}{\includegraphics{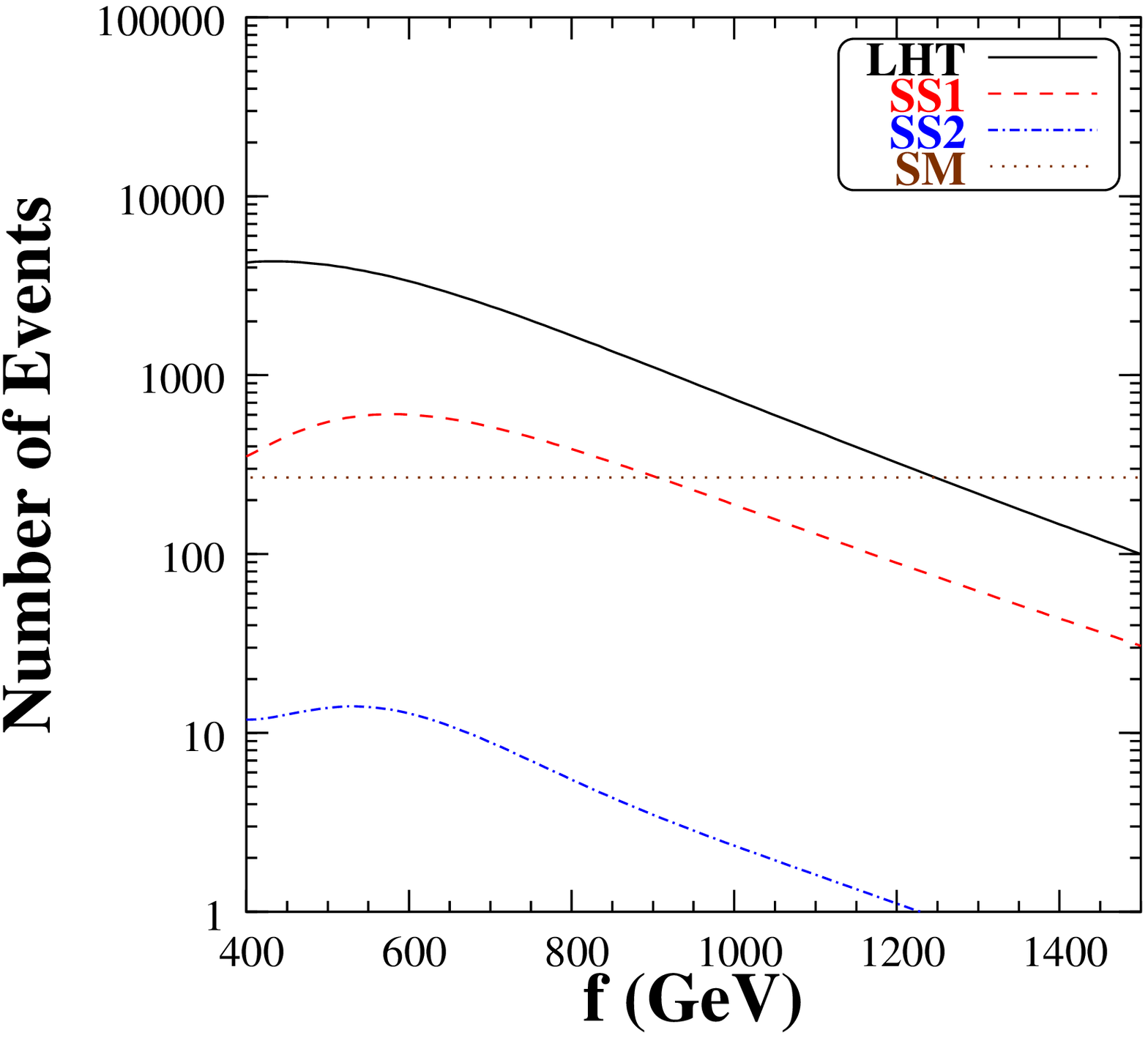}}
     \resizebox{75mm}{!}{\includegraphics{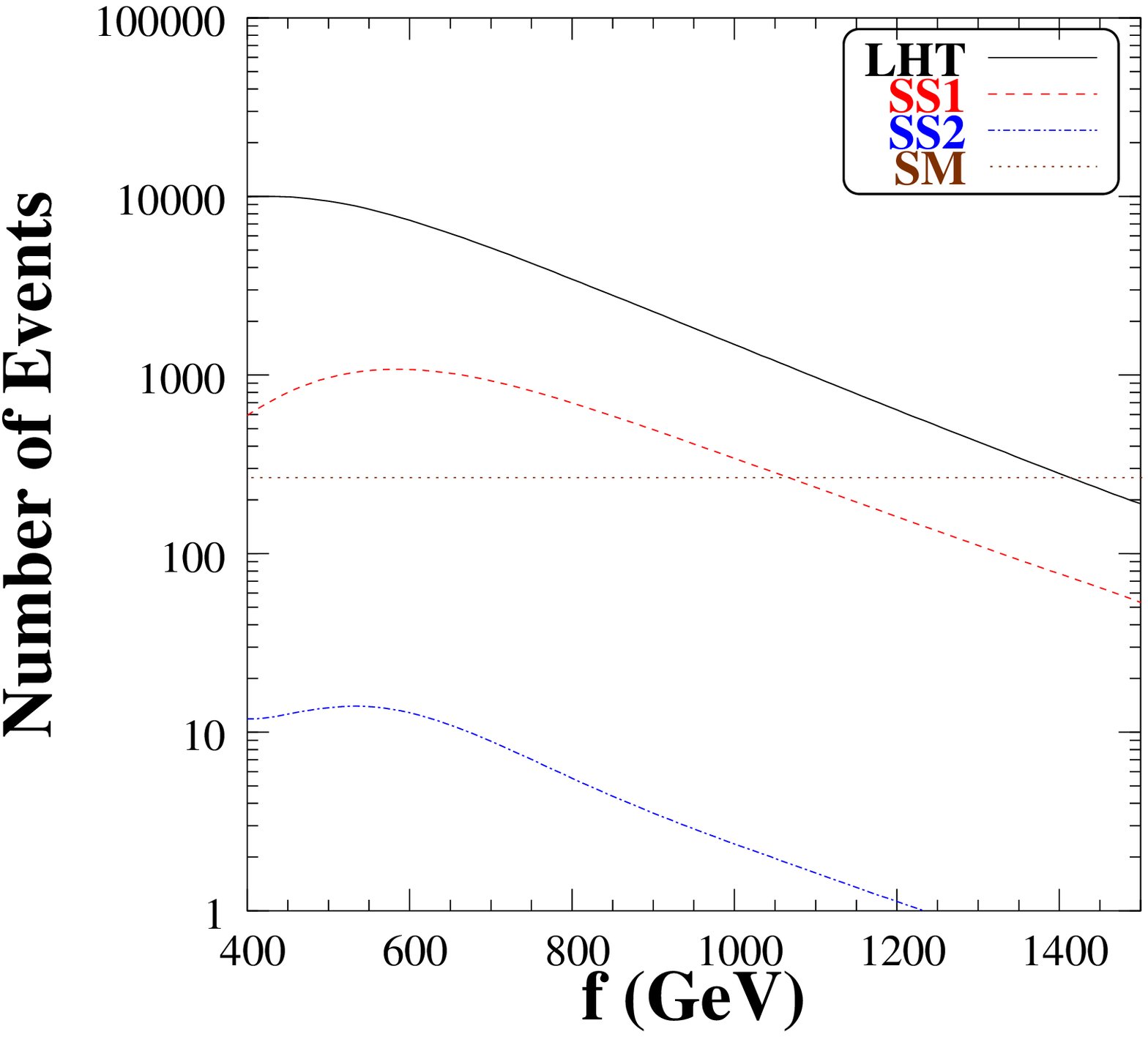}}
      \caption{$3l + E_T{\!\!\!\!\!\!/\ }$ event rates at the LHC after all the cuts imposed for LHT, SUSY scenarios SS1and SS2 with $\kappa_l=0.4$ for $\kappa_q = 1$ (left panel) and $\kappa_q =
	1.5$ (right panel) and the SM for an integrated luminosity of 
	$300~{\rm fb}^{-1}$. This plot is taken from the Ref.~\cite{ours}} 
 \label{rates1}
  \end{figure}

From Table~\ref{mass_table}, it is clear that with the aforementioned setup, all the SM partners can be matched within $10-20$ GeV mass difference.
For our analysis, we have chosen two set of $k_q =1$ and 1.5. Our choices of $k_l$ is such that in one case 
($k_l =.4$) heavy leptons are lighter than the heavy gauge bosons while for the other case ($k_l = 1$), the situation is reversed. Figure~\ref{rates1} we present variation of number of trilepton events
against the scale $f$ for LHT, SS1 and SS2 corresponding to the different level of cuts as mentioned earlier.
It is clear from Figure~\ref{rates1} that the LHT trilepton event rates remain higher after the cuts in
comparison to SS1 and SS2. This is primarily because of the larger
cross-sections for the LHT. The SS2 rates are further suppressed in comparison
to SS1 because of the small branching fractions for the leptonic decays of Higgsino like $\chione$ and $\chitwo$.

Clearly, the event rates are quite suppressed for cases with $k_l = 1$ which is due to the fact that now unlike the case with $k_l = .4$,  heavy leptonic decay modes does of $W_H$ and $Z_H$ does not contribute to the trileptonic signal.

%

\section{Summary and Conclusion}
In this talk, we have discussed the following:

\begin{itemize}
\item
\mbox{}{LHT trilepton events can be distinguished, at least at the
  $6\sigma$ level, from either SUSY scenario (SS1 and SS2) even for matching mass spectra for $\kappa_l < .44$}.
\item
For $\kappa_l >.44$ the distinction is not possible as the trilepton rates in LHT and SUSY are too low compared toSM background.
\item
We also noted that though a LHT-SUSY discrimination is possible for an integrated luminosity of 30~fb$^{-1}$,
 the information on the mass spectrum might not be sufficient at that stage of the LHC.
\end{itemize}

\begin{theacknowledgments}
This talk is based on works done in Ref.~\cite{ours} with A K Datta, P Dey, B Mukhopadhyaya and A Nyffeler. Fundings to attend this conference is provided by a DOE grant under contract number DE-FG02-01ER41155.
\end{theacknowledgments}

\end{document}